\documentclass[reprint,superscriptaddress,preprintnumbers,nofootinbib,amsmath,amssymb,aps]{revtex4-2}

\usepackage{graphicx}
\usepackage{dcolumn}
\usepackage{bm}

\usepackage{natbib}
\usepackage{booktabs}
\usepackage{enumitem}
\usepackage{amssymb}    
\usepackage{color}         
\usepackage{graphicx}     
\usepackage{mathrsfs} 
\usepackage{hyperref} 
\hypersetup{colorlinks=true,linkcolor=blue,citecolor=blue}
\usepackage{ upgreek } 
\usepackage{color,xcolor,fancybox,epsf,rotating,colordvi}

\newcommand{\TeV}{~\rm TeV}

\newcommand{\abm}{{~\rm ab}^{-1}}

\begin{document}


\title{Investigating higgsino dark matter in the semi-constrained NMSSM}
\author{Kun Wang}
\email[]{kwang@usst.edu.cn}
\affiliation{College of Science, University of Shanghai for Science and Technology,  Shanghai 200093,  China}

\author{Jingya Zhu}
\email[]{zhujy@henu.edu.cn} 
\affiliation{School of Physics and Electronics, Henan University,  Kaifeng 475004, China}


\date{\today}

\begin{abstract}
In this study, we explore the characteristics of higgsino-dominated dark matter (DM) within the semi-constrained Next-to-Minimal Supersymmetric Standard Model (scNMSSM), covering a mass range from hundreds of GeV to several TeV. 
We carefully analyzed the parameter space under existing theoretical and experimental constraints to confirm the viability of higgsino-dominated lightest supersymmetric particles (LSPs) with masses between 100 GeV and 4 TeV.
Our study examines various DM annihilation mechanisms, emphasizing the significant role of coannihilation with the next-to-lightest supersymmetric particle (NLSP), which includes other higgsino-dominated particles such as $\tilde{\chi}^{0}_2$ and $\tilde{\chi}^{\pm}_1$. We categorize the annihilation processes into three main classes: $\tilde{\chi}_1^{\pm}$ coannihilation, Higgs funnel annihilation, and $\tilde{\tau}_1$ coannihilation, each combines interactions with $\tilde{\chi}_1^{\pm}$.
Our results indicate that achieving the correct relic density in heavier higgsino LSPs requires a combination of coannihilation and Higgs funnel mechanisms. 
We also assess the potential of future experiments, such as XENONnT, LUX-ZEPLIN (LZ), PandaX-xT, and the Cherenkov Telescope Array (CTA), to probe these DM scenarios through direct and indirect detection.
In particular, future spin-independent DM detection can cover all samples with the correct DM relic density for $\mu \gtrsim 1300$ GeV.
Furthermore, future colliders like the International Linear Collider (ILC) and the Compact Linear Collider (CLIC) are found to exceed the detection capabilities of current hadron colliders, especially for higher mass NLSPs. 
Notably, CLIC at 3000 GeV is anticipated to thoroughly investigate all samples with insufficient DM relic density for $\mu \lesssim 1300$ GeV.
\end{abstract}

\maketitle
\newpage


\section{\label{sec:Introduction}Introduction}

Dark matter (DM) is a crucial yet elusive component of the Universe, accounting for $26.8\%$ of its total mass and energy. 
Despite its invisibility and the lack of direct detection, the presence of DM is strongly supported by observational evidence such as galaxy rotation curves, gravitational lensing, and anisotropies in the cosmic microwave background (CMB) \cite{Jungman:1995df,Bertone:2004pz,Bertone:2016nfn}. 
Among the proposed DM candidates, weakly interacting massive particles (WIMPs) are particularly noteworthy. 
They are consistent with the predictions of the Lambda Cold Dark Matter ($\Lambda$CDM) model and can have masses ranging from the MeV to TeV scale \cite{Boehm:2002yz, Boehm:2003bt, Murayama:2009nj, Hambye:2009fg}. 
The theoretical framework suggests that WIMPs were once in thermal equilibrium in the early universe and that their current abundance can be explained by the freeze-out mechanism\cite{Bernstein:1985th,Srednicki:1988ce}. 
This mechanism is crucial not only for understanding WIMP relic densities, but also for explaining other processes in the early universe, such as big bang nucleosynthesis and recombination\cite{Planck:2015fie}.
However, the continued lack of direct detection has led to broader investigations of WIMP properties, focusing on their potential annihilation channels and various detection methods.

The role of R-parity in supersymmetry (SUSY) is crucial since it ensures the stability of the lightest supersymmetric particle (LSP), making it a potential DM candidate. 
In the Minimal Supersymmetric Standard Model (MSSM), the LSP is usually a neutralino, which can be a bino ($\tilde{B}$), wino ($\tilde{W}$), or higgsino ($\tilde{H}_u$ or $\tilde{H}_d$) \cite{Gunion:1984yn,Gunion:1986nh,Haber:1984rc,Djouadi:2005gj}.
These four types of neutralino as LSPs have received lots of attention \cite{Barman:2024xlc,Harz:2023llw,Bisal:2023iip,Barman:2022jdg,Iwamoto:2021aaf,Cox:2021nbo,Arai:2020qxe,Ahmed:2020lua,Delgado:2020url,Borschensky:2018zmq,Araz:2018uyi,Abe:2018qlw,Kar:2017oer,Duan:2017ucw,Drees:2017xed,Hebbar:2017fit,Bringmann:2017sko,Belanger:2017vpq,Chakraborti:2017dpu}. 
For bino-dominated LSPs, the main problem is the risk of excessive DM. 
To avoid this by increasing the annihilation rate in the early universe, the preferred strategies include coannihilation mechanisms with wino, slepton, or stop \cite{Bagnaschi:2017tru,He:2023lgi}. 
If the LSP is not the only source of DM, a higgsino-dominated LSP in the 100-300 GeV mass range could still contribute to DM, since the DM-nucleon scattering cross section is significantly reduced due to the lower relic density \cite{Abdughani:2017dqs,Zhao:2022pnv}.
For a bino-higgsino mixed LSP, the efficient scattering with nucleons makes it difficult to meet the current direct detection limits. 
Overall, the lack of definitive results from DM detection experiments tends to constrain the parameter space, favoring a bino-dominated LSP.

In the Next-to-Minimal Supersymmetric Standard Model (NMSSM), which extends the MSSM by adding a singlet, the requirement for the 125 GeV Higgs mass puts less pressure on the stops, thus relaxing the tight constraints from DM experiments \cite{Baum:2017enm}. 
This modification introduces an additional singlino in the neutralino sector, drawing significant attention to the DM sector of the NMSSM \cite{Meng:2024lmi,Li:2023kbf,Heng:2023wqf,Binjonaid:2023rtc,Heng:2023vxf,Datta:2022bvg,Domingo:2022pde,Almarashi:2022iol,Tang:2022pxh,Chatterjee:2022pxf,Ahmed:2022jlo,Li:2021bka,Cao:2021tuh,Lopez-Fogliani:2021qpq,Zhou:2021pit,Cao:2021ljw,Abdallah:2020yag,Barman:2020vzm,Abdallah:2019znp,Cao:2019aam,Wang:2019biy,Cao:2012im}. 
Unlike in the MSSM, a bino-dominated DM candidate can co-annihilate with a singlino-dominated neutralino to achieve the measured abundance. 
In the NMSSM, DM is likely to reach the observed abundance primarily through two mechanisms: coannihilation with Higgsino-dominated electroweakinos (EWkinos), or through resonant annihilations facilitated by singlet-dominated CP-even or CP-odd Higgs bosons\cite{Griest:1988ma}. 
In the General NMSSM, singlino-like DM could achieve the observed relic abundance through different channels, such as $\chi^0_1\chi^0_1 \to h h$, with the theory significantly favoring singlino-dominated DM over bino-like DM over a wider range of parameters\cite{Meng:2024lmi}.

In our previous work\cite{Wang:2020xta,Wang:2020dtb,Wang:2020tap}, we focused on the funnel annihilations of light DM within the semi-constrained NMSSM (scNMSSM), which relaxes the Higgs masses at the GUT scale, also referred to as the NMSSM with non-universal Higgs mass.
We found that the light DM in the scNMSSM is typically singlino-dominated, with four possible funnel annihilation mechanisms allowing the singlino-dominated LSP to get the right relic density\cite{Wang:2020tap,Wang:2020dtb}. 
Furthermore, we found that highly singlino-dominated DM can reach sufficient relic density, but both the higgsino asymmetry and the spin-dependent cross section remain small. 
Conversely, when there is a sizeable higgsino component in the light DM, the relic density is always low\cite{Wang:2020xta}.
Therefore, in this work, we focus on higgsino dominated DM in the scNMSSM, aiming to determine if it has the correct DM relic density and to study its detection and annihilation processes.
We will examine the relic density of DM to ensure it agrees with cosmological observations and analyze the potential for direct and indirect detection based on current experimental limits. 
Additionally, we will explore future collider detection prospects and specific annihilation channels that could affect the detectability of Higgsino-dominated LSP.

The structure of this paper is organized as follows.
In Sec.~\ref{sec:model}, we provide a brief overview of the neutralino and chargino part of the scNMSSM.
In Sec.~\ref{sec:res}, we discuss our calculation methodology and provide a comprehensive analysis of the obtained results. 
In Sec.~\ref{sec:con}, we conclude by summarizing the main results and their implications for future researches.

\section{The semi-constrained NMSSM}
\label{sec:model}

The NMSSM extends the MSSM by adding a singlet superfield $\hat{S}$. The superpotential is modified to
\begin{equation}
    W_{\mathrm{NMSSM}} = W_{\mathrm{MSSM}}^{\mu=0} + \lambda \hat{S} \hat{H}_u \cdot \hat{H}_d + \frac{\kappa}{3} \hat{S}^3 \,,
\end{equation}
where $W_{\mathrm{MSSM}}^{\mu=0}$ represents the MSSM superpotential without the $\mu$-term. The singlet scalar's vacuum expectation value (VEV), $v_s$, generates the $\mu$-term dynamically
\begin{equation}
    \mu = \lambda v_s \,.
\end{equation}
The NMSSM also includes specific soft SUSY breaking terms:
\begin{align}
    -\mathcal{L}_{\mathrm{NMSSM}}^{\mathrm{soft}}  = 
    & -\mathcal{L}_{\mathrm{MSSM}}^{\mathrm{soft}}|_{\mu=0} + m_{S}^{2} | S |^{2} \nonumber \\
    &  + \lambda A_{\lambda} S H_{u} \cdot H_{d} + \frac{\kappa}{3} A_{\kappa} S^{3} + \mathrm{h.c.} \,.
\end{align}
where $H_u$ and $H_d$ are the Higgs doublet scalars, $A_{\lambda}$ and $A_{\kappa}$ are trilinear couplings, and $m_S$ is the singlet scalar soft mass.

In the scNMSSM, the Higgs sector can deviate from universality at the GUT scale, allowing the soft masses $m_{H_u}^2$, $m_{H_d}^2$, and $m_S^2$ to differ from $M_0^2 + \mu^2$. The trilinear couplings $A_{\lambda}$ and $A_{\kappa}$ can also can also be independent from $A_0$. Thus, the scNMSSM parameter space is defined by nine parameters:
\begin{equation}
    \lambda, \, \kappa, \, \tan\beta \equiv \frac{v_u}{v_d}, \, \mu, \, A_{\lambda}, \, A_{\kappa}, \, A_0, \, M_0 \,  M_{1/2}, \,
\end{equation}
where $M_{0}$ and $M_{1/2}$ are the universal sfermion and gaugino masses, and $A_{0}$ is the universal trilinear coupling in the sfermion sector. 

The electroweakino sector within the NMSSM is predicted to include five neutralinos and two pairs of charginos. 
The five neutralinos are composed of the following eigenstates $\tilde{B}$, $\tilde{W}_{3}$, $\tilde{H}_{d}^{0}$, $\tilde{H}_{u}^{0}$, and $\tilde{S}$. 
The mass matrix is
\begin{align}
M_{\tilde{\chi}^{0}}  & =  \nonumber \\ 
& \left(\begin{array}{ccccc}
M_{1} & 0 & -c_\beta s_W m_{Z} & s_\beta s_W m_{Z} & 0 \\
  & M_{2} & c_\beta c_W m_{z} & -s_\beta c_W m_{Z} & 0 \\
  &   & 0 & -\mu & -\lambda v_{d} \\
  &   &   & 0 & -\lambda v_{u} \\
  &   &   &   & 2 \kappa v_{s}
\end{array}\right)
\end{align}
where $s_\beta$, $c_\beta$, $s_W$, and $c_W$ represent $\sin \beta$, $\cos \beta$, $\sin \theta_W$, and $\cos \theta_W$, respectively.
And $M_1$ and $M_2$ are the bino and wino masses, respectively, evolved from $M_{1/2}$ through RGEs to the SUSY scale. 
The higgsino mass is $\mu$, and the singlino mass is $2 \kappa v_{s}$. 
Diagonalizing this matrix, one can obtain the five neutralino mass eigenstates, $\tilde{\chi}^{0}_i~(i=1...5)$.
The lightest neutralino is usually the LSP in the NMSSM, making it a perfect DM candidate.

The two pairs of charginos are composed of the following eigenstates $\tilde{H}^{\pm}$ and $\tilde{W}^{\pm}$. 
The mass matrix is
\begin{align}
M_{\tilde{\chi}^{\pm}} & = \left(\begin{array}{cc}
M_{2} & \sqrt{2} s_\beta m_{W} \\
\sqrt{2} c_\beta m_{W} & \mu
\end{array}\right)
\end{align}
where $M_2$ is the wino mass as mentioned above, and the charginos can be the mixture of the charged higgsinos and winos.
Diagonalizing this matrix, we can get two physical states in the chargino sector, typically denoted as $\tilde{\chi}^{\pm}_i~(i=1,2)$.
The lighter chargino, $\tilde{\chi}^{\pm}_1$, is dominated by either charged higgsinos or winos, depending on whether $\mu$ or $M_2$ is smaller.

The NMSSM Higgs sector includes three CP-even Higgs bosons ($H_i$, $i=1,2,3$), two CP-odd Higgs bosons ($A_i$, $i=1,2$), and a pair of charged Higgs bosons $H^{\pm}$. 
Of the CP-even Higgs bosons, one is the SM-like Higgs $H_{\rm SM}$, one is another doublet-dominated $H_D$, and one is singlet-dominated $H_S$
Similarly, of the two CP-odd Higgs bosons, one is doublet-dominated $A_D$ and the other is singlet-dominated $A_S$.

The coupling of the lightest neutralino $\tilde{\chi}^{0}_1$ to the Higgs basis states is given by \cite{Baum:2017enm}
\begin{align}
C_{H_{\rm SM} \tilde{\chi}^{0}_1 \tilde{\chi}^{0}_1 }   
&=\sqrt{2} \lambda N_{15}\left(N_{13} s_{\beta}+N_{14} c_{\beta}\right)  \nonumber \\
&+ \left(g_{1} N_{11}-g_{2} N_{12}\right)\left(N_{13} c_{\beta}-N_{14} s_{\beta}\right), \\
C_{H_D \tilde{\chi}^{0}_1 \tilde{\chi}^{0}_1 }   
&=\sqrt{2} \lambda N_{15}\left(N_{13} c_{\beta}-N_{14} s_{\beta}\right)  \nonumber \\
&-\left(g_{1} N_{11}-g_{2} N_{12}\right)\left(N_{13} s_{\beta}+N_{14} c_{\beta}\right), \\
C_{H_S \tilde{\chi}^{0}_1 \tilde{\chi}^{0}_1 } &=\sqrt{2}\left[\lambda N_{13} N_{14}-\kappa N_{15} N_{15}\right], \\
C_{A_S \tilde{\chi}^{0}_1 \tilde{\chi}^{0}_1}&=-i\sqrt{2}\left[\lambda N_{13} N_{14}-\kappa N_{15} N_{15}\right], \\
C_{A_D \tilde{\chi}^{0}_1 \tilde{\chi}^{0}_1 }   
&=i \big[ \sqrt{2} \lambda N_{15}\left(N_{13} c_{\beta}+N_{14} s_{\beta}\right) \nonumber \\ 
&- \left(g_{1} N_{11}-g_{2} N_{12}\right)\left(N_{13} s_{\beta}-N_{14} c_{\beta}\right)\big]
\end{align}
where $N_{ij}$ are obtained from the diagonalization of the neutralino mass matrix, and $C_{X \tilde{\chi}^{0}_1 \tilde{\chi}^{0}_1}$ denotes the couplings of the Higgs basis states ($H_{\rm SM}/H_D/H_S/A_D/A_S$) to the LSP $\tilde{\chi}^0_1$.

In the scNMSSM, the gaugino masses are unified at the GUT scale, resulting in the mass ratios at the SUSY scale $M_1 : M_2 : M_3 = 1 : 2 : 6$. Therefore, the mass of the wino ($M_2$) is larger than that of the bino ($M_1$), i.e., $M_2 > M_1$.
A higgsino-dominated LSP is quantified by $|N_{13}|^2 + |N_{14}|^2 > 0.5$ and has a mass close to $\mu$. 
This condition is met when $M_1$ and $2 \kappa v_{s}$ are all larger than $\mu$.
In this case, the NLSP (next-to-lightest supersymmetric particle) is another higgsino-dominated neutralino $\tilde{\chi}^{0}_2$ or a higgsino-dominated chargino $\tilde{\chi}^{\pm}_1$. 
Since both are higgsino-dominated, the mass of the NLSP is also close to $\mu$. 
Therefore, to achieve the correct relic density, the LSP can annihilate not only through funnel annihilation but also through coannihilation with the NLSP or other sparticles with similar masses.

When examining DM characteristics, the DM relic density of the LSP $\tilde{\chi}^{0}_1$ is critical. It is calculated by solving the Boltzmann equation for the number density $n$ of the LSP:
\begin{align}
    \frac{dn}{dt} = -3Hn - \langle v\sigma_{\text{eff}} \rangle (n^2 - n_{\text{eq}}^2),
\end{align}
where $H$ denotes the Hubble rate, representing the expansion rate of the universe; $n_{\text{eq}}$ is the equilibrium number density, and $\langle v\sigma_{\text{eff}} \rangle$ is the thermally averaged effective cross section for annihilation and coannihilation processes.

The relic density $\Omega h^2$ is then calculated from the equilibrium solutions of this equation. It can be expressed through the simplified formula \cite{Young:2016ala}:
\begin{align}
    \Omega h^2 = \frac{m_{\tilde{\chi}^{0}_1} n_0 h^2}{\rho_c},
\end{align}
where $m_{\tilde{\chi}^{0}_1}$ is the mass of the LSP, $n_0$ is the current number density of the DM, and $\rho_c = {3H_0^2}/{8\pi G_N}$ represents the critical density of the Universe. 
In our study, we utilized the \textsf{micrOMEGAs} package to calculate the DM relic density and the scattering cross-section of DM with nucleons.

\section{Results and Discussions}
\label{sec:res}

In this study, we focus on the relic density and annihilation channels of higgsino-dominated DM in the scNMSSM, considering a wide mass range from GeV to TeV. 
We also explore the implications of direct and indirect detection experiments, as well as collider searches, on the properties and viability of higgsino-dominated DM.
We require the LSP to be higgsino-dominated, which is defined by
\begin{align}
 |N_{13}|^2 + |N_{14}|^2 > 0.5
\end{align}
Therefore, we consider the relevant parameter space in the scNMSSM as follows:
\begin{gather*}
0.0<\lambda<0.7, \quad  |\kappa|<0.7, \quad 1<\tan\beta <60 \, , \\
0.1<\mu< 10\TeV , \quad 0.0< M_0, \,\, M_{1/2} < 10\TeV \, , \\
|A_0|,\,\, |A_\lambda|,\,\, |A_\kappa| < 10 \TeV \, .
\end{gather*}

The parameter space scan are consistent with our previous work \cite{Wang:2021fsz}, using the \textsf{NMSSMTools-6.0.2} package \cite{Ellwanger:2004xm, Ellwanger:2005dv, Ellwanger:2006rn, Das:2011dg} to conduct the scans and calculate relevant properties.
Both theoretical and experimental constraints are considered: theoretical constraints include vacuum stability and absence of Landau poles \cite{Ellwanger:2004xm, Ellwanger:2005dv}; experimental constraints include flavor constraints \cite{Tanabashi:2018oca,  Aaij:2012nna, Lees:2012xj, Lees:2012ym}, a global fit of LHC Higgs data \cite{Aad:2019mbh, Sirunyan:2018koj, Khachatryan:2016vau,ATLAS:2022vkf,CMS:2022dwd,CMS:2018uag,ATLAS:2016neq}, DM relic density \cite{Hinshaw:2012aka, Ade:2013zuv}, direct and indirect DM searches \cite{Aprile:2018dbl, Aprile:2019dbj,Amole:2019fdf,PandaX:2022xas,PandaX-4T:2021bab,LZ:2022lsv}. 
Constraints on SUSY searches at the LHC and LEP are implemented using the \textsf{SModelS-v2.2.1} package \cite{Kraml:2013mwa, Ambrogi:2017neo, Ambrogi:2018ujg, Dutta:2018ioj, Buckley:2013jua, Sjostrand:2006za, Sjostrand:2014zea}, while constraints on searches for additional Higgs bosons and exotic decays of the SM-like Higgs are applied using the \textsf{HiggsBounds-5.5.0} package \cite{Bechtle:2015pma, Bechtle:2013wla, Bechtle:2013gu, Bechtle:2011sb, Bechtle:2008jh,ATLAS:2023tkt}.

For the samples that satisfy  the above  theoretical and experimental constraints, we observe the following properties.
The mass range of the lightest neutralino varies from 100 GeV to 4 TeV and is predominantly higgsino-dominated. 
The mass of the lightest neutralino, predominantly higgsino-dominated, ranges from 100 GeV to 4 TeV. Similarly, the mass of the lightest chargino, $\tilde{\chi}_{1}^{\pm}$, also spans from 100 GeV to 4 TeV, mirroring that of the lightest neutralino.
This similarity arises because both the lightest neutralino and the lightest chargino are higgsino-dominated, resulting in their masses being approximately equal to $\mu$. Since $\mu < M_1$, this sets an upper limit on their masses.
In the Higgs sector, $h_1$ is identified as the 125 GeV SM-like Higgs. $h_2$ and $h_3$ are classified as heavy CP-even Higgs bosons, while $a_1$ and $a_2$ are heavy CP-odd Higgs bosons.

\begin{figure*}[!htbp]
\centering
\includegraphics[width=1\textwidth]{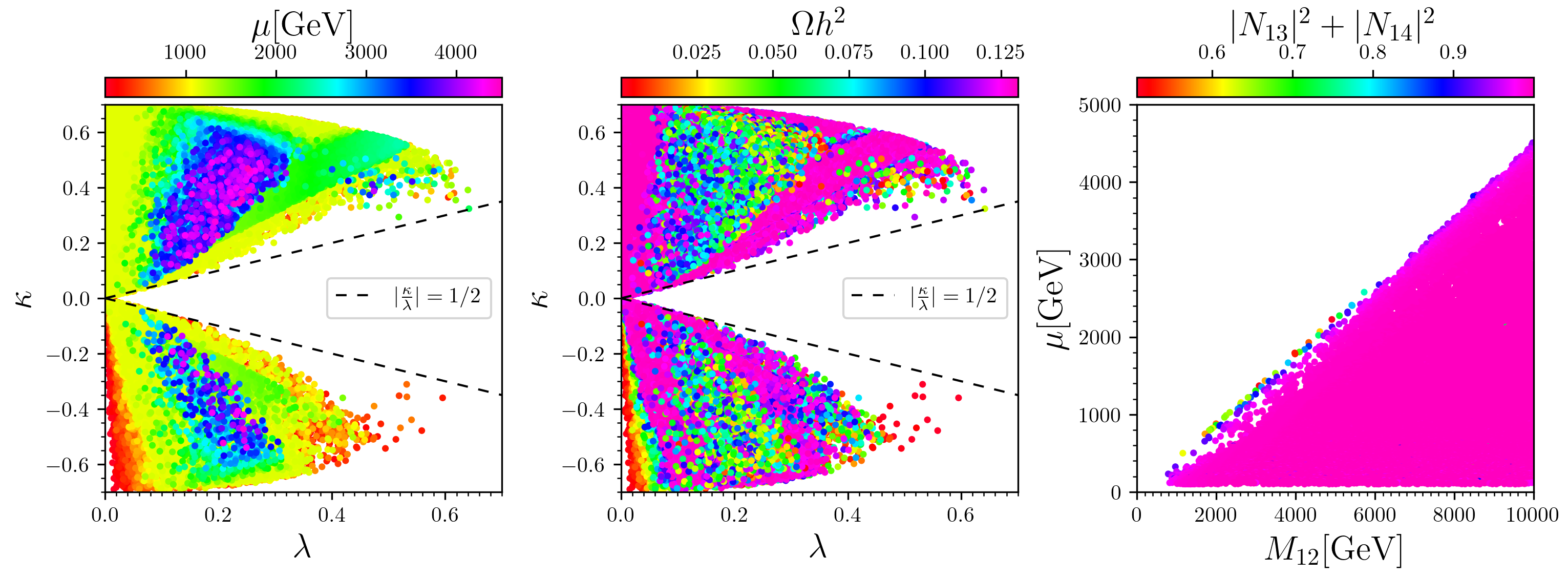}
\vspace{-0.6cm}
\caption{
Surviving samples are shown in the planes of $\kappa$ versus $\lambda$ (left and middle) and $\mu$ versus $M_{1/2}$ (right). 
From left to right, the colors represent $\mu$, the DM relic density $\Omega h^2$, and the higgsino component in the LSP $|N_{13}|^2 + |N_{14}|^2$, respectively. 
Samples with larger values of $\mu$ are plotted on top of those with smaller values.
}
\label{fig:1}
\end{figure*}

In Fig.~\ref{fig:1}, we show the DM relic density and higgsino component of LSP in the scNMSSM.
we can see following key points from this figure:
\begin{itemize}
\item In the left and middle panels, the surviving samples are shown on the $\kappa$ versus $\lambda$ plane. For the higgsino-dominated LSP, where the higgsino is the lightest neutralino and the singlino has a larger mass, the following relationship is given
\begin{align}
2 \kappa v_{s} = 2 \kappa \frac{\mu}{\lambda} > \mu.
\end{align}
Therefore, all surviving samples are in the region $|\kappa/\lambda| > 1/2$.

\item In the left panel, the color indicates $\mu$. We observe that $\mu \gtrsim 3$ TeV is restricted to a small region. This is because, as the higgsino mass increases, the DM relic density becomes excessively large. The coannihilation rate slows down, necessitating funnel annihilation processes to reduce the DM density.

\item In the middle panel, the color indicates the DM relic density $\Omega h^2$. We find that in the region with larger $\mu$, which corresponds to a heavy higgsino LSP, there are few samples with the correct relic density. This is because heavy higgsino LSPs require both coannihilation and funnel annihilation to achieve the correct relic density.
Samples with $\mu \approx 1000$ GeV appear to have the right DM relic density, likely because they achieve the correct relic density primarily through coannihilation. In contrast, samples with $\mu \lesssim 1000$ GeV do not have the correct relic density.

\item In the right panel, the surviving samples are displayed on the $\mu$ versus $M_{1/2}$ plane, with the color indicating the higgsino component in the LSP, $|N_{13}|^2 + |N_{14}|^2$. We can observe that almost all the samples are highly higgsino-dominated, with $|N_{13}|^2 + |N_{14}|^2 \gtrsim 0.9$.
Additionally, $M_{1/2}$ constrains the upper bound of the higgsino LSP. Since $\mu < M_1$ and the gaugino masses follow the ratio $M_1 : M_2 : M_3 = 1 : 2 : 6$, the bino mass $M_1$ is proportional to $M_{1/2}$.

\end{itemize}

\begin{figure*}[!tbp]
\centering
\includegraphics[width=1\textwidth]{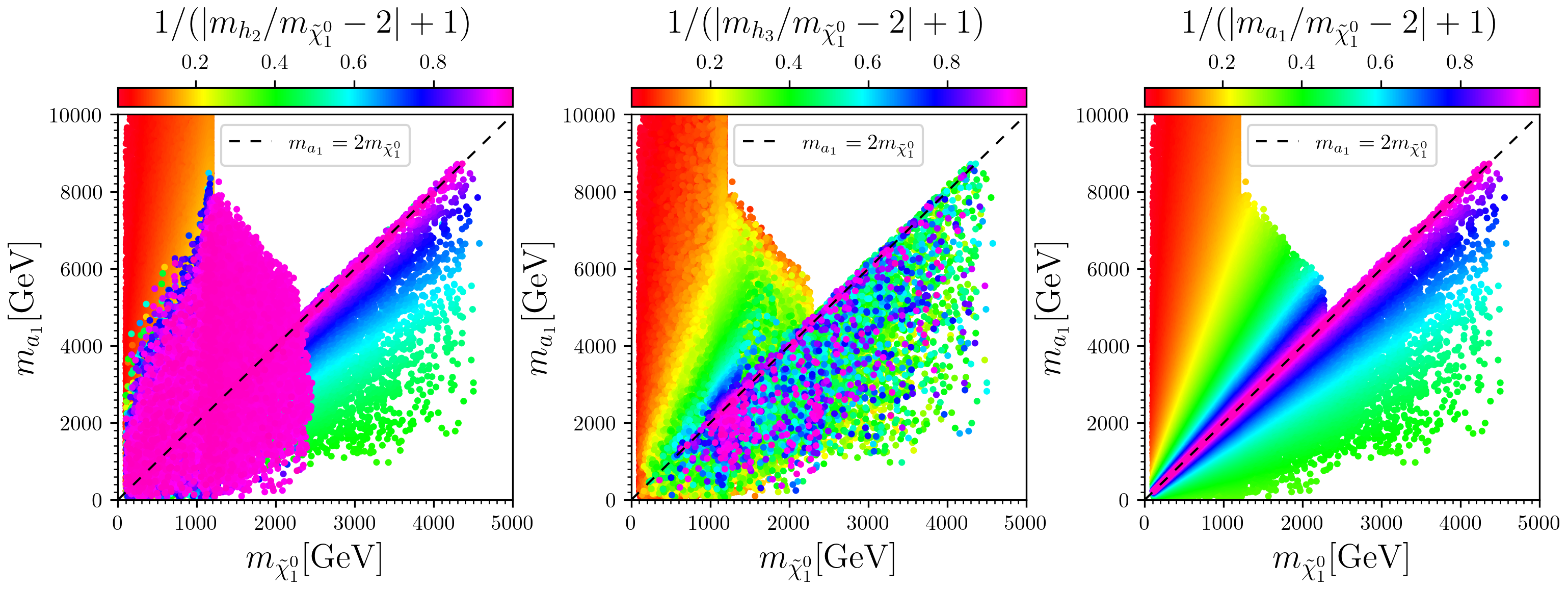}
\includegraphics[width=1\textwidth]{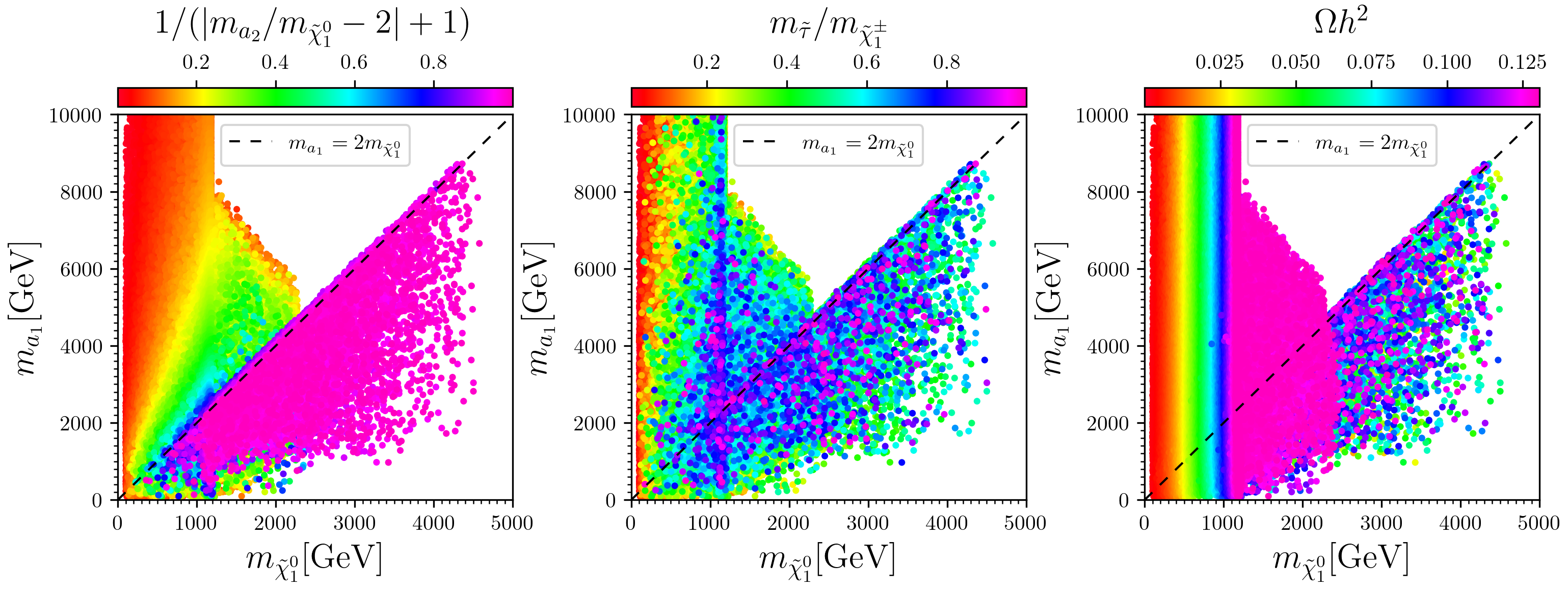}
\caption{
The surviving samples are displayed on the planes of the light CP-odd Higgs boson mass $m_{a_1}$ versus the LSP mass $m_{\tilde{\chi}^{0}_1}$. 
Colors represent relevant Higgs-LSP mass degeneracies (upper panels and lower left panel), stau-LSP mass degeneracies (lower middle panel) and DM relic density $\Omega h^2$ (lower right panel).
The color of the Higgs-LSP mass degeneracies $1/(|m_{H/A}/m_{\tilde{\chi}_1^{0}}-2|+1)$ approaching one indicates that Higgs masses tend to be $2 m_{\tilde{\chi}_{1}^{0}}$. 
Similarly, the stau-LSP mass degeneracies $m_{\tilde{\tau}}/m_{\tilde{\chi}_1^{\pm}}$ approaching one indicate that stau masses tend to be $m_{\tilde{\chi}_{1}^{0}}$.
Colors closer to zero indicate larger mass degeneracies.
In the upper panels, lower left and middle panels, samples with smaller mass degeneracies are plotted on top of those with smaller values, while in the lower right panels, samples with larger DM relic density $\Omega h^2$ are plotted on top.
}
\label{fig:2}
\end{figure*}

In Fig.~\ref{fig:2}, the surviving samples are displayed on the planes of the light CP-odd Higgs boson mass $m_{a_1}$ versus the LSP mass $m_{\tilde{\chi}^{0}_1}$. 
Colors represent relevant mass degeneracies (upper, lower left, and lower middle panels) and DM relic density $\Omega h^2$ (lower right panels).
The color of the Higgs-LSP mass degeneracies $1/(|m_{H/A}/m_{\tilde{\chi}_1^{0}}-2|+1)$ approaching one indicates that Higgs masses tend to be $2 m_{\tilde{\chi}_{1}^{0}}$. 
Similarly, the stau-LSP mass degeneracies $m_{\tilde{\tau}}/m_{\tilde{\chi}_1^{\pm}}$ approaching one indicate that stau masses tend to be $m_{\tilde{\chi}_{1}^{0}}$.
Colors closer to zero indicate highly mass degeneracies.
The mass degeneracies of different annihilation mechanisms are based on the following properties \cite{Buchmueller:2014yva,Bagnaschi:2015eha}:
\begin{align}
\label{eq:co1} 
\tilde{\chi}_{1}^{ \pm} \text { coannihilation :} \quad &\left( \frac{m_{\tilde{\chi}_{1}^{ \pm}}}{m_{\tilde{\chi}_{1}^{0}}}-1 \right) < 0.1 \, ,\\
\label{eq:co2}
\tilde{\tau}_{1} \text { coannihilation :} \quad &\left( \frac{m_{\tilde{\tau}_{1}}}{m_{\tilde{\chi}_{1}^{0}}} - 1 \right) < 0.15, \, ,\\
\label{eq:co3}
\tilde{t}_{1} \text { coannihilation :} \quad &\left( \frac{m_{\tilde{t}_{1}}}{m_{\tilde{\chi}_{1}^{0}}}\right)-1 < 0.2 \, , \\
\label{eq:co4}
A / H \text { funnel : } \quad &\left| \frac{m_{A/H}}{m_{\tilde{\chi}_{1}^{0}}}-2 \right| <0.4 \, .
\end{align}

We observe that the coannihilation mechanism requires the NLSP mass to be very close to the LSP mass. 
For instance, chargino $\tilde{\chi}_{1}^{\pm}$, stau $\tilde{\tau}_{1}$, and stop $\tilde{t}_{1}$ coannihilation occur when their masses are very close to the LSP mass, leading to the corresponding coannihilation. 
Specifically, for the higgsino-LSP scenario in the scNMSSM, we find that all samples satisfy $\tilde{\chi}_{1}^{\pm}$ coannihilation, with a small fraction satisfying $\tilde{\tau}_{1}$ coannihilation, and no samples satisfying $\tilde{t}_{1}$ coannihilation. 

The funnel annihilation mechanism requires the CP-even or CP-odd Higgs $H/A$ mass to be close to twice the LSP mass. 
For surviving samples, $h_1$ is the 125 GeV SM-like Higgs, and the minimum LSP mass is larger than 100 GeV, so a higgsino-dominated LSP does not undergo $h_1$ funnel annihilation. 
Only $h_2$, $h_3$, $a_1$, and $a_2$ funnel annihilation mechanisms are present.
In the scNMSSM framework, our analysis shows that there are samples with a higgsino-dominated LSP that achieve the correct DM relic density with no need for funnel annihilation.
However, for heavier higgsino-dominated LSP, funnel annihilation is necessary to achieve the correct DM relic density.

In Fig.~\ref{fig:2}, we used $1/(|m_{H/A}/m_{\tilde{\chi}_1^{0}}-2|+1)$ to represent the Higgs-LSP mass degeneracy. 
This expression approximate to one when the condition for the A/H funnel mechanism is satisfied, i.e., $m_{\tilde{\chi}_1^{0}} \approx 2m_{H/A}$, and it approaches zero in the opposite case.
We can observe the following from this figure:
\begin{itemize}
\item For the surviving samples, all samples satisfy the $\tilde{\chi}_{1}^{\pm}$ coannihilation condition. 
This is because the masses of both the higgsino-dominated LSP and the chargino $\tilde{\chi}^{\pm}_1$ are approximately equal to $\mu$.

\item In the upper left panel, we can see that $h_2$ funnel samples seem to occupy two distinct regions on the $m_{a_1}$ versus $m_{\tilde{\chi}^{0}_1}$ panel. 
One region is concentrated where $m_{\tilde{\chi}^{0}_1} \lesssim 2400 \rm GeV$ and $m_{a_1} \lesssim 8000 \rm GeV$; the other region is concentrated around $m_{a_1} = 2m_{\tilde{\chi}^{0}_1}$.

\item In the upper middle panel, we can see that $h_3$ funnel samples are relatively few and seem to be distributed in the region where $m_{a_1} < 2m_{\tilde{\chi}^{0}_1}$.
In the upper right panel, we can see that $a_1$ funnel samples seem to be distributed around the line $m_{a_1} = 2m_{\tilde{\chi}^{0}_1}$.
In the lower left panel, we can see that $a_2$ funnel samples are relatively more abundant and seem to be distributed in the region where $m_{a_1} < 2m_{\tilde{\chi}^{0}_1}$.

\item In the lower middle panel, the distribution of samples satisfying $\tilde{\tau}_1$ coannihilation is analyzed. It is observed that these samples are dispersed throughout the plane, indicating that $\tilde{\tau}_1$ coannihilation occurs across all examined regions.

\item In the lower right panel, there seems to be a distinct region where the higgsino-dominated LSP mass is less than 1300 GeV. In this region, the DM relic density is insufficient, but it increases with mass until it reaches around 1300 GeV, where the DM relic density becomes sufficient.
This is because all samples satisfy the $\tilde{\chi}_{1}^{\pm}$ coannihilation condition. 
When the LSP mass is low, the coannihilation process is more efficient, leading to easier annihilation of DM and resulting in an insufficient DM relic density.
As the LSP mass increases, coannihilation decreases, causing the DM relic density to rise. 
When the LSP mass exceeds 1300 GeV, $\tilde{\chi}_{1}^{\pm}$ coannihilation becomes less efficient, leading to an excess of DM. 
At this point, additional funnel annihilation mechanisms are needed to assist in DM annihilation.

\end{itemize}

\begin{figure*}[!htbp]
\centering
\includegraphics[width=1.0\textwidth]{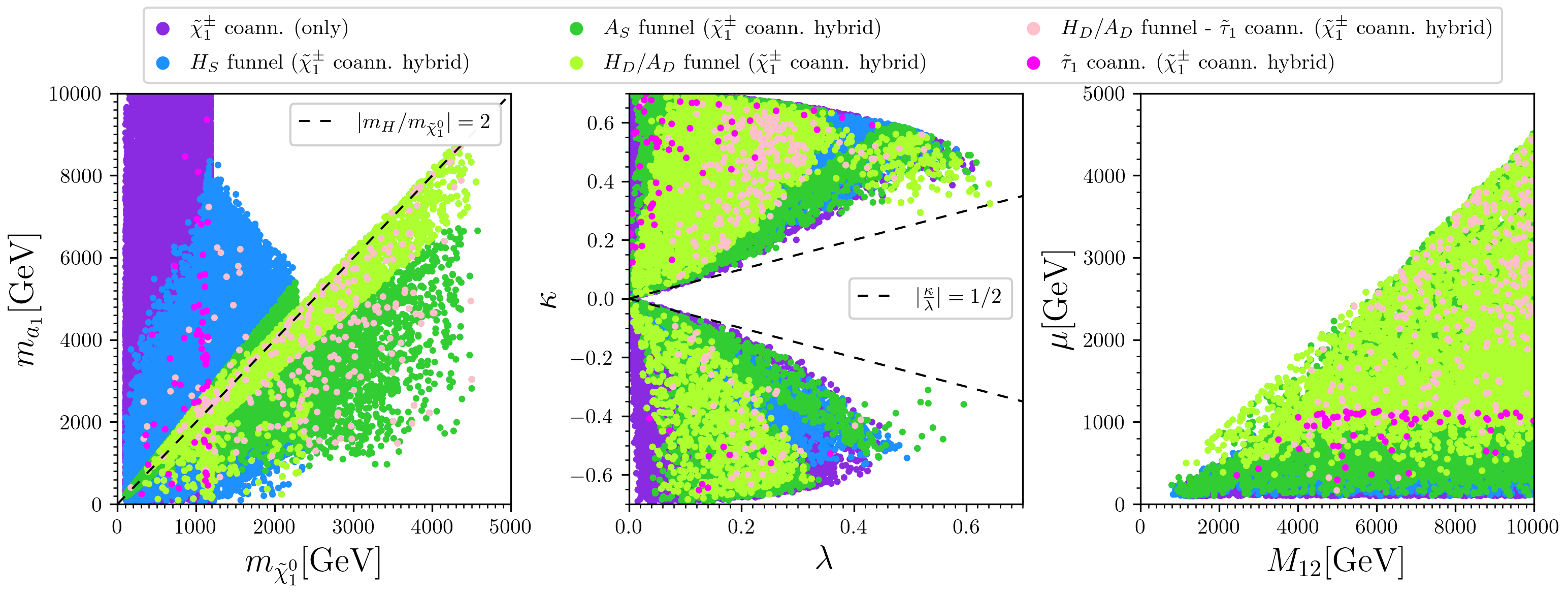}
\caption{
Surviving samples are shown in the planes of the light CP-odd Higgs boson mass $m_{a_1}$ versus the LSP mass $m_{\tilde{\chi}^{0}_1}$ (left), $\kappa$ versus $\lambda$ (middle) and $\mu$ versus $M_{1/2}$ (right). 
Each color represents a different DM annihilation mechanism. 
All samples satisfy the $\tilde{\chi}_1^{\pm}$ annihilation criteria and are classified according to the DM annihilation mechanisms they meet as follows: only the $\tilde{\chi}_1^{\pm}$ coannihilation (purple), singlet-dominated CP-even Higgs funnel annihilation (blue), singlet-dominated CP-odd Higgs funnel annihilation (dark green), doublet-dominated Higgs funnel annihilation (light green), doublet-dominated Higgs funnel annihilation with $\tilde{\tau}_1$ coannihilation (pink), and $\tilde{\tau}_1$ coannihilation (magenta).
}
\label{fig:3}
\end{figure*}

In Fig.~\ref{fig:3}, the surviving samples are displayed on the planes of the light CP-odd Higgs boson mass $m_{a_1}$ versus the LSP mass $m_{\tilde{\chi}^{0}_1}$ (left), $\kappa$ versus $\lambda$ (middle) and $\mu$ versus $M_{1/2}$ (right). 
Each color represents a different DM annihilation mechanism. 
Since all surviving samples have a higgsino-dominated LSP, and the light chargino $\tilde{\chi}^{\pm}_1$ is also higgsino-dominated, each sample inherently satisfies the $\tilde{\chi}_1^{\pm}$ annihilation criterion.
Therefore, except for the points representing only $\tilde{\chi}_1^{\pm}$ coannihilation (purple), all other points correspond to hybrid annihilation mechanisms.
We have categorized the samples into six classes based on the different DM annihilation mechanisms they satisfy:
\begin{itemize}
    \item $\tilde{\chi}_1^{\pm}$ coann. (only): solely the $\tilde{\chi}_1^{\pm}$ coannihilation;
    \item $H_S$ funnel ($\tilde{\chi}_1^{\pm}$ coann. hybrid): singlet-dominated CP-even Higgs funnel annihilation combined with $\tilde{\chi}_1^{\pm}$ coannihilation;
    \item $A_S$ funnel ($\tilde{\chi}_1^{\pm}$ coann. hybrid): singlet-dominated CP-odd Higgs funnel annihilation combined with $\tilde{\chi}_1^{\pm}$ coannihilation;
    \item $H_D/A_D$ funnel ($\tilde{\chi}_1^{\pm}$ coann. hybrid): doublet-dominated Higgs funnel annihilation combined with $\tilde{\chi}_1^{\pm}$ coannihilation;
    \item $H_D/A_D$ funnel - $\tilde{\tau}_1$ coann. ($\tilde{\chi}_1^{\pm}$ coann. hybrid): doublet-dominated Higgs funnel annihilation combined with $\tilde{\tau}_1$ coannihilation and $\tilde{\chi}_1^{\pm}$ coannihilation;
    \item $\tilde{\tau}_1$ coann. ($\tilde{\chi}_1^{\pm}$ coann. hybrid): $\tilde{\tau}_1$ coannihilation combined with $\tilde{\chi}_1^{\pm}$ coannihilation.
\end{itemize}
The classification of these samples is based on the criteria described in the Eq.~\eqref{eq:co1}--\eqref{eq:co4}.
When comparing Fig.~\ref{fig:3} with Fig.~\ref{fig:1} and Fig.~\ref{fig:2}, the following conclusions can be drawn:
\begin{itemize}
\item In the left panel, it is observed that the blue samples, which represent only $\tilde{\chi}_1^{\pm}$ coannihilation, correspond to an LSP mass below 1300 GeV. 
This annihilation channel constrains the upper mass limit of the LSP due to the DM relic density requirements. 
As shown in the lower right panel of Fig.~\ref{fig:2}, the DM relic density increases with larger LSP masses. 
This is because the efficiency of $\tilde{\chi}_1^{\pm}$ coannihilation decreases as the mass increases. 
Consequently, larger LSP masses, if solely relying on $\tilde{\chi}_1^{\pm}$ coannihilation, would result in a DM relic density that exceeds the experimental upper limits.

\item In the left panel, for samples with an LSP mass below 1300 GeV, a fraction also falls into the category of $\tilde{\tau}_1$ coannihilation combined with $\tilde{\chi}_1^{\pm}$ coannihilation. 
It is important to note that the primary mechanism for these parameter points remains $\tilde{\chi}_1^{\pm}$ coannihilation. 
For LSP masses above 1300 GeV, achieving the correct DM relic density requires the use of Higgs funnel annihilation mechanisms. 
Depending on the proportion of singlet versus doublet components in the Higgs, their distribution varies.
Blue samples, satisfying the conditions for $H_S$ funnel annihilation, typically have LSP masses below 2500 GeV. 
In contrast, samples satisfying the conditions for $A_S$ funnel annihilation (dark green) and $H_D/A_D$ funnel annihilation (light green) can span any LSP mass range. 
Pink samples, which involve funnel annihilation combined with $\tilde{\tau}_1$ coannihilation, also cover any LSP mass. 
However, the main contributors to the DM relic density in these samples are still $\tilde{\chi}_1^{\pm}$ coannihilation and Higgs funnel annihilation.

\item In the middle panel, combined with the middle panel of Fig.~\ref{fig:1}, we can observe that in the region with a large $\mu$ value, corresponding to a heavy higgsino-dominated LSP, $A_S$ funnel annihilation (dark green) and $H_D/A_D$ funnel annihilation (light green) are possible. 
Additionally, there is a small presence of Higgs funnel annihilation combined with $\tilde{\tau}_1$ coannihilation (pink). 
Only a few samples in this region reach the correct DM relic density.

\item In the left panel, within the region where $\mu$ is relatively low, we can observe different probabilities of Higgs funnel annihilation. 
For instance, in regions where $\mu$ is less than 600 GeV, $A_S$ funnel annihilation (dark green) is more likely to occur. 
When $\mu$ decreases to below 200 GeV, $H_S$ funnel annihilation (blue) becomes more common.

\end{itemize}

\begin{figure*}[!htbp]
\centering
\includegraphics[width=1.0\textwidth]{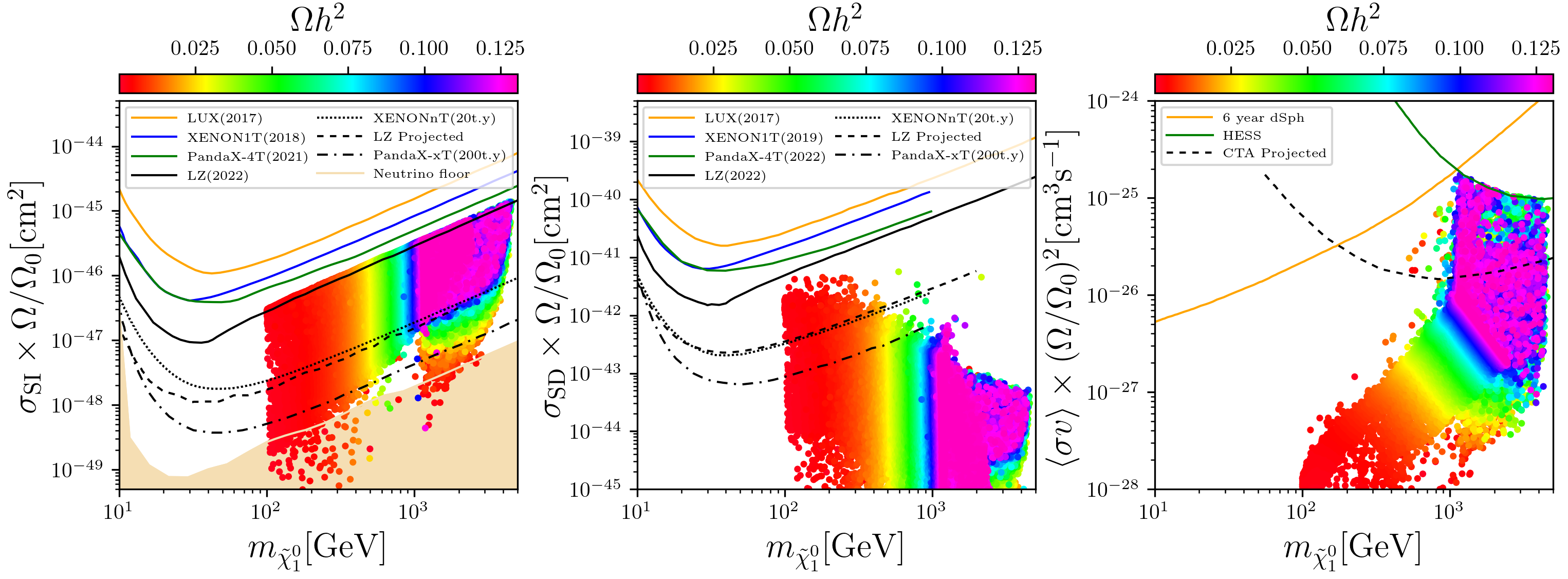}
\includegraphics[width=1.0\textwidth]{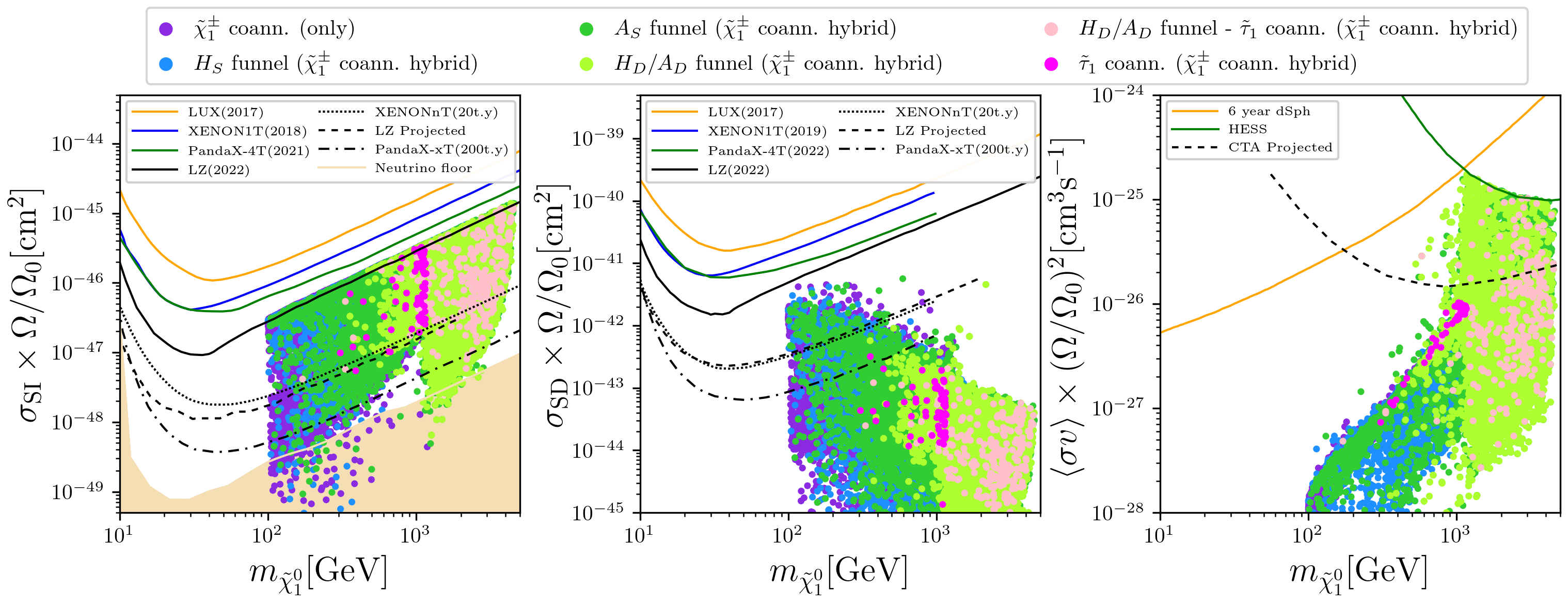}
\caption{
Surviving samples are displayed across the planes of the rescaled spin-independent DM-nucleon cross-section $\sigma_{\rm SI} \times \Omega/\Omega_0$ (left), rescaled spin-dependent DM-nucleon cross-section $\sigma_{\rm SD} \times \Omega/\Omega_0$ (middle), and rescaled thermally averaged DM annihilation cross section times velocity $\langle \sigma v \rangle  \times (\Omega / \Omega_0)^2$ (right) versus LSP mass $m_{\tilde{\chi}^{0}_1}$, where $\Omega_0 = 0.1187$.
In the upper panels, colors indicate the DM relic density $\Omega h^2$. 
In the lower panels, colors represent different DM annihilation mechanisms, as in Fig.~\ref{fig:3}. 
Samples with higher DM relic densities $\Omega h^2$ are plotted above those with lower values in the upper panels.
In the left panels, the orange, blue, green, and black solid curves indicate the spin-independent (SI) DM-nucleon cross-section detection limits of LUX (2017) \cite{LUX:2016ggv}, XENON1T (2018) \cite{XENON:2018voc}, PandaX-4T (2021) \cite{PandaX-4T:2021bab}, and LZ (2022) \cite{LZ:2022lsv}, respectively. 
The black dotted, dashed, and dot-dashed curves indicate the future detection limits of XENONnT (20t.y) \cite{XENON:2020kmp}, LUX-ZEPLIN (LZ) projected \cite{LZ:2018qzl}, and PandaX-xT (200t.y) \cite{PandaX:2024oxq}, respectively. The neutrino floor \cite{Billard:2013qya} is indicated by the orange shaded region.
In the middle panels, the orange, blue, green, and black solid curves indicate the spin-dependent (SD) DM-nucleon cross-section detection limits of LUX (2017) \cite{LUX:2017ree}, XENON1T (2019) \cite{XENON:2019rxp}, PandaX-4T (2022) \cite{PandaX:2022xas}, and LZ (2022) \cite{LZ:2022lsv}, respectively. 
The black dotted, dashed, and dot-dashed curves indicate the future detection limits of XENONnT (20t.y) \cite{XENON:2020kmp}, LZ projected \cite{LZ:2018qzl}, and PandaX-xT (200t.y) \cite{PandaX:2024oxq}, respectively.
In the right panels, the orange and green solid curves indicate the upper limits on the pair-annihilation rate from the Milky Way dSphs based on 6 years of Fermi Large Area Telescope data \cite{Fermi-LAT:2015att} and 112 hours of observations of the Galactic Center with HESS \cite{HESS:2011zpk}, respectively. 
The black dashed curves indicate projected sensitivity from the Cherenkov Telescope Array (CTA) on the pair-annihilation rate \cite{Workman:2022ynf}.
}
\label{fig:4}
\end{figure*}

In Fig.~\ref{fig:4}, surviving samples are displayed on the planes of the rescaled spin-independent (SI) DM-nucleon cross-section  $\sigma_{\rm SI} \times \Omega/\Omega_0$ (left), rescaled spin-dependent (SD) DM-nucleon cross-section $\sigma_{\rm SD} \times \Omega/\Omega_0$ (middle), and rescaled thermally averaged DM annihilation cross section times velocity $\langle \sigma v \rangle  \times (\Omega / \Omega_0)^2$ (right) versus LSP mass $m_{\tilde{\chi}^{0}_1}$, where $\Omega_0 = 0.1187$.
The colors indicate the DM relic density $\Omega h^2$ (upper panels), and different DM annihilation mechanisms (lower panels). 
We used rescaled SI/SD DM-nucleon cross-section and rescaled DM pair annihilation rate because the LSP may not constitute all of the DM. 
The following conclusions can be drawn from this figure.
\begin{itemize}
\item In the upper left and middle panels, samples more easily escape SD DM-nucleon cross-section constraints, whereas the SI DM-nucleon cross-section constraint is significantly stricter. 
This is because the higgsino-dominated LSP has a very small higgsino asymmetry, denoted as $|N_{13}^2 - N_{14}^2|$. Since the SD cross-section is proportional to the square of this asymmetry \cite{Wang:2020xta}, it remains low, enabling samples to easily escape experimental SD constraints.

\item In the upper left panels, future experiments such as XENONnT (20t.y) and LUX-ZEPLIN (LZ) (1000 day) are expected to cover all samples with the correct DM relic density, while the upcoming PandaX-xT (200t.y) will cover nearly all samples, approaching the neutrino floor.
In the lower left panels, XENONnT (20t.y) and the LZ (1000 days) are expected to cover nearly all samples with $\tilde{\tau}_1$ coannihilation, and the upcoming PandaX-xT (200t.y) will cover different Higgs funnel annihilations.
Only samples with insufficient DM relic density are likely to escape future experimental constraints.

\item In the middle panels, future SD DM-nucleon cross section constraints from XENONnT (20t.y), LZ (1000 days), and PandaX-xT (200t.y) will only be able to cover a small fraction of samples with insufficient DM relic density, including parts of the $\tilde{\chi}_1^{\pm}$ coannihilation (only) and $H_S$ and $A_S$ funnel annihilation samples. 
In the right panels, future DM annihilation rate constraints from the Cherenkov Telescope Array (CTA) projections will include some samples with sufficient DM relic density, including parts of the $A_S$ and $H_D/A_D$ funnel annihilation samples.

\end{itemize}

\begin{figure*}[!htbp]
\centering
\includegraphics[width=1.0\textwidth]{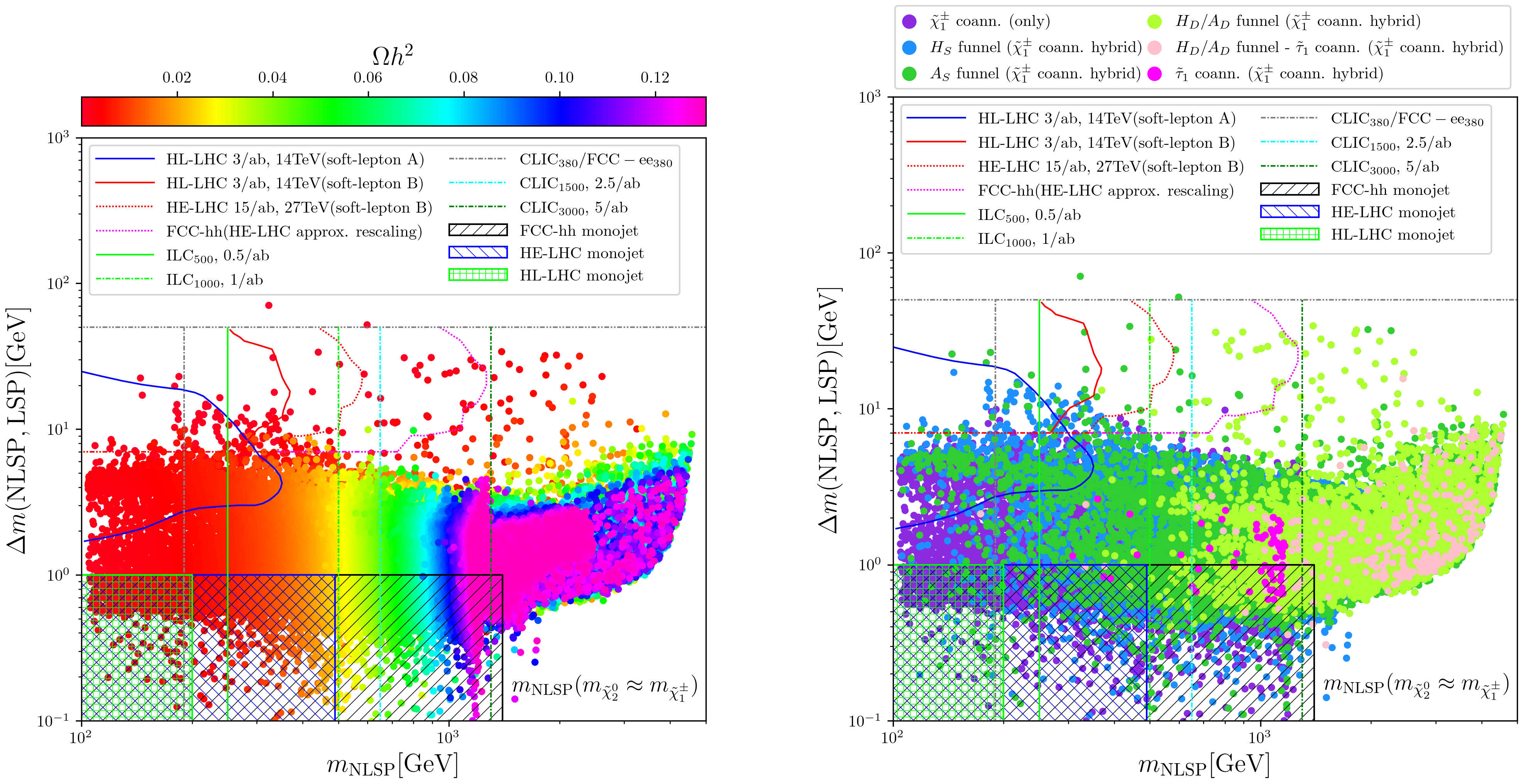}
\caption{
Surviving samples are displayed in the planes of mass difference between NLSP and LSP, $\Delta m (\rm NLSP, LSP)$, versus NLSP mass $m_{\rm NLSP}$, where higgsino-dominated charginos and next-to-lightest neutralinos as NLSP have nearly identical masses, $m_{\tilde{\chi}_2^{0}}\approx m_{\tilde{\chi}_1^{\pm}}$.
The left panel uses colors to represent the DM relic density $\Omega h^2$, with samples having larger values plotted above those with smaller values.
The right panel uses colors to illustrate various DM annihilation mechanisms. 
Detection limits for higgsino-dominated electroweak processes by future detectors are taken from Fig.8.10 in Ref.\cite{EuropeanStrategyforParticlePhysicsPreparatoryGroup:2019qin}.
The blue and red curves represent detection limits at the HL-LHC with $3\abm$ at 14 TeV using soft leptons; red and magenta dotted curves depict the HE-LHC with $15\abm$ at 27 TeV and the FCC-hh with $30\abm$ at 100 TeV \cite{ATLAS:2018jjf}, respectively. 
The light green solid and dot-dashed curves are for the International Linear Collider (ILC) at $0.5\abm$ at 500 GeV and $1\abm$ at 1000 GeV \cite{Berggren:2020tle}, respectively. 
The gray, light blue, and green dot-dashed curves represent Compact Linear Collider (CLIC) at $1\abm$ at 380 GeV, $2.5\abm$ at 1500 GeV, and $5\abm$ at 3000 GeV \cite{EuropeanStrategyforParticlePhysicsPreparatoryGroup:2019qin}, respectively. 
The black, blue, and light green slash-hatched regions indicate mono-jet searches at the FCC-hh, HE-LHC, and HL-LHC \cite{CidVidal:2018eel,Golling:2016gvc}, respectively.
}
\label{fig:5}
\end{figure*}

In Fig.~\ref{fig:5}, surviving samples are displayed on the planes of the mass difference between NLSP and LSP, $\Delta m (\rm NLSP, LSP)$, versus the NLSP mass $m_{\rm NLSP}$. 
Here, higgsino-dominated charginos and next-to-lightest neutralinos, serving as NLSP, have nearly equal masses, $m_{\rm NLSP} = m_{\tilde{\chi}2^{0}} \approx m{\tilde{\chi}_1^{\pm}}$, indicating a compressed EWkino spectrum. 
Detection limits for higgsino-dominated electroweak processes by future detectors are derived from Fig.8.10 in Ref.\cite{EuropeanStrategyforParticlePhysicsPreparatoryGroup:2019qin}. 
For HL-LHC, HE-LHC, and FCC-hh, the constraints are based on a search involving two soft leptons in the production of $\tilde{\chi}_2^{0} \tilde{\chi}_1^{\pm}$, $\tilde{\chi}_2^{0} \tilde{\chi}_1^{0}$, and $\tilde{\chi}_1^{\pm} \tilde{\chi}_1^{\pm}$ \cite{ATLAS:2018jjf}. 
For the International Linear Collider (ILC), the constraints come from searches for $\tilde{\chi}_1^{\pm}$ through disappearing track analyses in a higgsino-dominated LSP scenario \cite{Berggren:2020tle}. 
For a very close mass difference $\Delta m (\rm NLSP, LSP) < 1 \text{GeV}$, monojet searches have been considered at the FCC-hh, HE-LHC, and HL-LHC \cite{CidVidal:2018eel, Golling:2016gvc}.
Several conclusions can be drawn from this figure.
\begin{itemize}

\item In the left panel, samples with insufficient DM relic density and NLSP mass below 350 GeV can be probed at HL-LHC for mass splittings $\Delta m (\rm NLSP, LSP)$ between 1.6 and 50 GeV. 
The NLSP detection limit at HE-LHC is about 1.5 times higher than at HL-LHC, while FCC-hh projections show NLSP detection limits up to 1 TeV, with sensitivity dependent on mass splittings $\Delta m (\rm NLSP, LSP)$. 
The sensitivity of lepton colliders is independent of $\Delta m (\rm NLSP, LSP)$. 
The $\rm ILC_{500}$ and $\rm ILC_{1000}$ can detect NLSP masses up to 250 GeV and 500 GeV, respectively, while $\rm CLIC_{1500}$ and $\rm CLIC_{3000}$ can reach 650 GeV and 1300 GeV.
Monojet searches at hadron colliders can extend the search for small $\Delta m (\rm NLSP, LSP)$ scenarios. 
For $\Delta m (\rm NLSP, LSP) < 1 \text{GeV}$, the higgsino monojet search could reach an NLSP mass of 200 GeV, 490 GeV, and 1400 GeV at the 14 TeV HL-LHC, 27 TeV HE-LHC, and 100 TeV FCC-hh colliders, respectively.

\item In the right panel, we can see that samples with the Higgs funnel annihilation mechanism are difficult to probe with hadron colliders and require lepton colliders. 
The lepton colliders $\rm ILC_{500}$ and $\rm ILC_{1000}$ can cover some samples with $H_S$ and $A_S$ funnel annihilation mechanisms, while $\rm CLIC_{3000}$ can cover some samples with $H_D/A_D$ funnel annihilation mechanisms, as well as some samples with $\tilde{\tau}_{1}$ coannihilation.

\end{itemize}

\begin{table*}[]
\centering
\caption{\label{tab:1}
Five Benchmark Points for Surviving Samples, where the relative contributions of channels to $\Omega h^2$ represent the percentage contributions of different DM annihilation channels to the DM relic density.
}
\setlength{\tabcolsep}{6.0pt}
\begin{tabular}{@{\hspace{2pt}}cccccc@{\hspace{2pt}}}
\toprule
                                              & P1       & P2       & P3       & P4       & P5       \\ \midrule
$\lambda$                                     &$1.64 \times 10^{-4}$ & $2.52 \times 10^{-3}$ & $3.98 \times 10^{-1}$ & $9.58 \times 10^{-2}$ & $1.58 \times 10^{-1}$ \\

$\kappa$                                      & 0.26     & 0.21     & 0.54     & -0.19    & 0.54     \\
$\tan\beta$                                   & 5.7      & 10.5     & 7.1      & 26.1     & 30.6     \\
$\mu \rm [GeV]$                               & 302      & 1103     & 1348     & 3800     & 823      \\
$M_0 \rm [GeV]$                               & 6820     & 9303     & 4917     & 6156     & 218      \\
$M_{12} \rm [GeV]$                            & 8024     & 9964     & 6983     & 9141     & 6765     \\
$A_0 \rm [GeV]$                               & -5023    & -2643    & 6675     & -3336    & -2497    \\
$A_\lambda \rm [GeV]$                         & 9900     & -4528    & 7246     & 7035     & -4788    \\
$A_\kappa \rm [GeV]$                          & -3427    & -8386    & -6337    & 3064     & -7811    \\
$m_{h_1} \rm [GeV]$                           & 125      & 126      & 124      & 126      & 126      \\
$m_{h_2} \rm [GeV]$                           & 28555    & 31225    & 2803     & 5590     & 4044     \\
$m_{h_3} \rm [GeV]$                           & 939301   & 183648   & 7865     & 14031    & 4815     \\
$m_{a_1} \rm [GeV]$                           & 28556    & 31225    & 4184     & 5589     & 4044     \\
$m_{a_2} \rm [GeV]$                           & 65007    & 46134    & 7865     & 7751     & 4915     \\
$\tilde{\chi}^{0}_1 \rm [GeV]$                & 316      & 1145     & 1385     & 3864     & 849      \\
$\tilde{\chi}^{0}_2 \rm [GeV]$                & 318      & 1146     & 1389     & 3868     & 851      \\
$\tilde{\chi}^{0}_3 \rm [GeV]$                & 3728     & 4689     & 3228     & 4288     & 3110     \\
$\tilde{\chi}^{0}_4 \rm [GeV]$                & 6767     & 8445     & 3784     & 7681     & 5607     \\
$\tilde{\chi}^{0}_5 \rm [GeV]$                & 1002573  & 193954   & 5842     & 14999    & 5753     \\
$\tilde{\chi}^{\pm}_1 \rm [GeV]$              & 317      & 1145     & 1387     & 3867     & 850      \\
$\tilde{\chi}^{\pm}_2 \rm [GeV]$              & 6767     & 8445     & 5842     & 7681     & 5607     \\
$\tilde{t}_1 \rm [GeV]$                       & 9214     & 12772    & 9658     & 12007    & 8338     \\
$\tilde{\tau}_1 \rm [GeV]$                    & 7332     & 9991     & 5365     & 6118     & 857      \\
$\Omega h^2$                                  & 0.01     & 0.12     & 0.07     & 0.12     & 0.09     \\
$\langle \sigma v \rangle \rm [cm^3 s^{-1}] $ & $9.67 \times 10^{-26}$ & $7.78 \times 10^{-27}$ & $5.25 \times 10^{-27}$ & $3.07 \times 10^{-26}$ & $1.41 \times 10^{-26}$ \\
$\sigma_{\rm SI} \rm [cm^2]$                  & $4.35 \times 10^{-47}$ & $2.71 \times 10^{-47}$ & $2.42 \times 10^{-46}$ & $4.12 \times 10^{-46}$ & $6.18 \times 10^{-47}$ \\
$\sigma_{\rm SD} \rm [cm^2]$                  & $1.71 \times 10^{-43}$ & $1.20 \times 10^{-44}$ & $8.26 \times 10^{-46}$ & $2.02 \times 10^{-44}$ & $4.62 \times 10^{-44}$  \\ \midrule
\begin{tabular}[c]{@{}c@{}} 
Relative \\ Contributions \\ of Channels \\ to $\Omega h^2$
\end{tabular} 
&  \begin{tabular}[c]{@{}c@{}}
$\tilde{\chi}_1^+ \tilde{\chi}^0_1 \to \text{SM36.2\%}$\\
$\tilde{\chi}_1^+ \tilde{\chi}^0_2 \to \text{SM23.5\%}$\\
$\tilde{\chi}_1^+ \tilde{\chi}_1^- \to \text{SM19.7\%}$\\
$\tilde{\chi}^0_1 \tilde{\chi}^0_2 \to \text{SM15.2\%}$\\
$\tilde{\chi}^0_1 \tilde{\chi}^0_1 \to \text{SM 4.0\%}$\\
$\tilde{\chi}^0_2 \tilde{\chi}^0_2 \to \text{SM 1.4\%}$
\end{tabular}        
& \begin{tabular}[c]{@{}l@{}}
$\tilde{\chi}_1^+ \tilde{\chi}^0_1 \to \text{SM31.7\%}$\\
$\tilde{\chi}_1^+ \tilde{\chi}^0_2 \to \text{SM27.4\%}$\\
$\tilde{\chi}_1^+ \tilde{\chi}_1^- \to \text{SM20.1\%}$\\
$\tilde{\chi}^0_1 \tilde{\chi}^0_2 \to \text{SM15.7\%}$\\
$\tilde{\chi}^0_1 \tilde{\chi}^0_1 \to \text{SM 2.9\%}$\\
$\tilde{\chi}^0_2 \tilde{\chi}^0_2 \to \text{SM 2.0\%}$
\end{tabular}         
&    \begin{tabular}[c]{@{}l@{}}
$\tilde{\chi}_1^+ \tilde{\chi}_1^- \to \text{SM 38.5\%}$\\
$\tilde{\chi}^0_1 \tilde{\chi}^0_1 \to \text{SM 19.6\%}$\\
$\tilde{\chi}^0_2 \tilde{\chi}^0_2 \to \text{SM 13.5\%}$\\
$\tilde{\chi}_1^+ \tilde{\chi}^0_1 \to \text{SM 12.7\%}$\\
$\tilde{\chi}_1^+ \tilde{\chi}^0_2 \to \text{SM  9.9\%}$\\
$\tilde{\chi}^0_1 \tilde{\chi}^0_2 \to \text{SM  6.0\%}$
\end{tabular}      
&  \begin{tabular}[c]{@{}l@{}}
$\tilde{\chi}_1^+ \tilde{\chi}_1^- \to \text{SM 45.0\%}$\\
$\tilde{\chi}^0_1 \tilde{\chi}^0_1 \to \text{SM 28.2\%}$\\
$\tilde{\chi}^0_2 \tilde{\chi}^0_2 \to \text{SM 19.9\%}$\\
$\tilde{\chi}_1^+ \tilde{\chi}^0_1 \to \text{SM  2.2\%}$\\
$\tilde{\chi}_1^+ \tilde{\chi}^0_2 \to \text{SM  1.8\%}$\\
$\tilde{\chi}^0_1 \tilde{\chi}^0_2 \to \text{SM  0.8\%}$
\end{tabular}        
&   \begin{tabular}[c]{@{}l@{}}
$\tilde{\chi}_1^+ \tilde{\chi}^0_1 \to   \text{SM 32.6\%}$\\
$\tilde{\chi}_1^+ \tilde{\chi}^0_2 \to   \text{SM 24.8\%}$\\
$\tilde{\chi}_1^+ \tilde{\chi}_1^- \to   \text{SM 19.5\%}$\\
$\tilde{\chi}^0_1 \tilde{\chi}^0_2 \to   \text{SM 15.3\%}$\\
$\tilde{\chi}^0_1 \tilde{\chi}^0_1 \to   \text{SM  3.8\%}$\\
$\tilde{\chi}^0_2 \tilde{\chi}^0_2 \to   \text{SM  1.7\%}$\\
$\tilde{\chi}_1^+ \tilde{\tau}_1 \to     \text{SM  0.8\%}$\\
$\tilde{\chi}^0_1 \tilde{\tau}_1 \to     \text{SM  0.4\%}$\\
$\tilde{\chi}^0_2 \tilde{\tau}_1 \to     \text{SM  0.3\%}$\\
$\tilde{\tau}_1 \bar{\tilde{\tau}}_1 \to \text{SM  0.1\%}$\\
$\tilde{\tau}_1 \tilde{\tau}_1 \to       \text{SM  0.1\%}$
\end{tabular}       \\ \bottomrule
\end{tabular}
\end{table*}

In Table~\ref{tab:1}, we present five benchmark points detailing the DM sector, where the relative contributions of channels to $\Omega h^2$ represent the percentage contributions of different DM annihilation channels to the DM relic density. Points 1 and 2 belong to the sole $\tilde{\chi}_1^{\pm}$ coannihilation; it is observed that a smaller LSP mass leads to an insufficient DM relic density in these cases. Points 3 and 4 are associated with Higgs funnel annihilation combined with $\tilde{\chi}_1^{\pm}$ coannihilation, showing that $\tilde{\chi}_1^{\pm}$ coannihilation still predominantly contributes to the DM relic density. Point 5 belongs to the $\tilde{\tau}_1$ coannihilation combined with $\tilde{\chi}_1^{\pm}$ coannihilation, where $\tilde{\chi}_1^{\pm}$ coannihilation remains the main contributor and $\tilde{\tau}_1$ coannihilation contributes only about $1.7\%$.

\section{Conclusion}
\label{sec:con}

In this study, we explore the higgsino-dominated DM relic density, annihilation mechanism, and detectability within the scNMSSM, covering a mass range from 100 GeV to over 4 TeV.
We start by scanning the parameter space with \textsf{NMSSMTools}, taking into account both theoretical and experimental constraints, such as vacuum stability, Higgs boson data, SUSY searches at the LHC, DM relic density, and direct and indirect DM searches.
To ensure the higgsino components are predominant in the LSP, we impose the condition $|N_{13}|^2 + |N_{14}|^2 > 0.5$, confirming all samples contain higgsino-dominated LSPs suitable as DM candidates. 
We confirm that $h_1$ is consistent with the 125 GeV SM-like Higgs, and the LSP mass is between 100 GeV and 4 TeV.

Further, we compute the DM relic density and explore various annihilation mechanisms. 
We assess the potential for both direct and indirect detection using existing experimental limits and consider the prospects for detection at future colliders. 
In this study, we investigated the characteristics and annihilation mechanisms of the lightest supersymmetric particle (LSP) and the next-to-lightest supersymmetric particle (NLSP) within a Higgsino-dominated framework. 
Our findings are summarized as follows:
\begin{itemize}
\item \textbf{Characterization of LSP and NLSP: } The mass of the Higgsino-dominated LSP is closely related to the $\mu$ parameter, and the NLSP is typically either another Higgsino-dominated neutralino, $\tilde{\chi}^{0}_2$, or a Higgsino-dominated chargino, $\tilde{\chi}^{\pm}_1$. Given their Higgsino domination, the NLSP mass similarly approximates the $\mu$ value.

\item \textbf{Annihilation Mechanisms: } The primary annihilation mechanism for the Higgsino-dominated LSP involves coannihilation with its NLSP counterpart, $\tilde{\chi}^{\pm}_1$. This study further identifies several annihilation pathways, including Higgs funnel annihilation mechanisms and $\tilde{\tau}_1$ coannihilation processes. We have categorized the particle samples into three main categories based on their respective DM annihilation mechanisms:
\begin{itemize}
    \item Sole $\tilde{\chi}_1^{\pm}$ coannihilation;
    \item Higgs funnel annihilation mechanisms involving $H_S$, $A_S$, and $H_D/A_D$, combined with $\tilde{\chi}_1^{\pm}$ coannihilation hybrids;
    \item $\tilde{\tau}_1$ coannihilation, which may or may not include Higgs funnel annihilation mechanisms, alongside $\tilde{\chi}_1^{\pm}$ coannihilation.
\end{itemize}

\item \textbf{DM Relic Density and Detection: } All examined samples satisfy the $\tilde{\chi}_1^{\pm}$ coannihilation criteria. The efficiency of the coannihilation mechanism decreases with increasing LSP mass, affecting the DM relic density. Particularly, samples with a larger $\mu$ (heavier LSP) exhibit lower probabilities of achieving the correct relic density without additional funnel annihilation mechanisms. The higgsino asymmetry ($|N{13}^2 - N_{14}^2|$) significantly influences the spin-dependent (SD) DM-nucleon cross-section, enabling many samples to evade experimental SD constraints.

\item \textbf{Future Experimental Prospects: } Upcoming experiments, such as XENONnT, LZ (1000 days), and PandaX-xT, aim to cover nearly all samples reach the correct DM relic density ($\mu \gtrsim 1300$GeV) through SI DM direct detection. 
In parallel, the Cherenkov Telescope Array (CTA) is expected to further scrutinize DM annihilation rates for indirect detection, particularly through $A_S$ and $H_D/A_D$ funnel mechanisms.
Additionally, the detection capabilities at lepton colliders surpass those at hadron colliders. The future ILC and CLIC are expected to detect samples with higher mass NLSPs effectively. Notably, the CLIC at 3000 GeV could probe all samples with insufficient DM relic density ($\mu \lesssim 1300$ GeV).

\end{itemize}

\begin{acknowledgments}
This work was supported by the National Natural Science Foundation of China (NNSFC) under
grant Nos. 12275066 and 11605123.
\end{acknowledgments}

\appendix





\bibliographystyle{apsrev4-1}
\bibliography{apssamp}

\end{document}